\documentclass[journal=jctcce, manuscript=article, layout=twocolumn, email=true]{achemso}
\setkeys{acs}{articletitle=true}
\usepackage[T1]{fontenc}
\usepackage[utf8]{inputenc}
\usepackage{amsfonts}
\usepackage{multirow,dcolumn}
\usepackage{booktabs}
\usepackage[version=3]{mhchem}
\usepackage{setspace} 
\usepackage{lmodern} 
\usepackage{adjustbox}
\usepackage{leftidx}
\usepackage{mathtools}
\usepackage{arydshln}
\usepackage{caption}
\usepackage{tabularx}
\usepackage{threeparttable}
\usepackage[dvipsnames]{xcolor}
\usepackage{blindtext} 
\usepackage{booktabs}
\usepackage[version=3]{mhchem}
\usepackage{lmodern}
\usepackage{natbib}
\usepackage{graphicx}
\usepackage[normalem]{ulem}
\usepackage{xcccom}
\usepackage{titlesec}
\titleformat{\section}
  {\normalfont\fontsize{12}{15}\bfseries}{\thesection}{1em}{}

\usepackage{braket}
\newcommand{\PP}{\mathcal{P}}
\newcommand{\paldus}{\texttt{Paldus}}

\title{F-12 density matrices and cumulants
  from the  explicitly connected coupled-cluster theory}

\author{Aleksandra M. Tucholska}
\email{aleksandra.tucholska@gmail.com}
\affiliation{Faculty of Chemistry, University of Warsaw, Pasteura 
  1, 02-093 Warsaw, Poland}
\author{Marcin Modrzejewski}
\affiliation{Faculty of Chemistry, University of Warsaw, Pasteura 
  1, 02-093 Warsaw, Poland}
  
\author{Robert Moszynski}
\affiliation{Faculty of Chemistry, University of Warsaw, Pasteura 
  1, 02-093 Warsaw, Poland}
\begin{document}
\begin{abstract}
We present the expansion to the expectation value coupled cluster theory (XCC) to the wavefunctions that include the inter electronic distances $r_{12}$ explicitly. We have extended our algebraic manipulation code \paldus to deal with the rems arising in the CC-F12 theory. We present the full working expressions for the one-electron density matrix (1RDM) and cumulant of the two-electron density matrix ($\lambda$-2RDM) in the framework of XCC-F12 theory. We analyze  the computational cost and discuss the possible approximations  the expressions.

\end{abstract}
\begin{center}
\maketitle
\end{center}
\section{Introduction}
For the computation of the molecular properties of small- and medium-sized systems the coupled cluster (CC) theory\cite{scuseria1987closed, bartlett1978many, bartlett2007coupled} 
is the leading \textit{ab initio} approach. CC method is size extensive and allows for  systematic approximation by including selected excitations.  Currently CC is routinely used for the computation of ground-state energies, molecular properties, excited states, etc.

Still, to obtain chemical accuracy ($< 1 $ kcal mol$^{-1}$) without including costly, higher excitations, one needs to address the incompleteness of the basis set that causes the well-known basis set error. It originates from the fact that  one-electron orbitals are used  to construct two-electron basis sets. It was known since 1957 Kato's discovery of the cusp condition\cite{kato1957eigenfunctions} that the inclusion of the inter electronic distance $r_{12}$ explicitly in the wave function might lead to the construction of an efficient wavefunction. 

The main obstacle of using such methods are the high-dimension integrals arising in the theory.
So far numerous approaches to deal with this problem have been proposed, form the direct evaluation of the high-dimension integrals\cite{wind2001efficient, wind2002second}, through expanding the correlation factor in terms of Gaussian Geminals\cite{persson1996accurate, may2004explicitly}, to the well known R12/F12 methods proposed by Kutzelnig\cite{kutzelnigg1985r, kutzelnigg1991wave} where through the insertion of the resolution of identity (RI) only two-electron integrals remain.
Numerous approaches have been developed to deal with this problem.  Among them the standard approximation idea (SA), proposed by Kutzelnig and Klopper, to introduce the resolution of identity (RI) to the integrals which allows for a reduction of the three- and four-electron integrals to two-electron terms. 

Although the SA simplified the integrals, the $r_{12}$ methods still required to use large basis sets.\cite{kutzelnigg1991wave, noga1997cc, rohse1993configuration}  This problem was addressed be Klopper and Samson\cite{klopper2002explicitly} by the introduction of ABS basis - additional basis set for the RI. Valeev\cite{valeev2004improving} proposed a robust modification to this approach called the complementary auxiliary basis set - CABS method 
which involves expansion in the orthogonal complement to the span of orbital basis set (OBS).
which will be utilized in this work.
The large CABS basis is used only for the RI terms and the normal orbital basis set is retained for the rest of the terms, making the $r_{12}$ methods feasible.

Within the standard approximation the CC-F12 theory was first presented by Noga and collaborators.\cite{noga1994coupled} The exponential form generates highly nonlinear, complicated expressions, therefore it is a common practice to further approximate the expressions for the amplitude, e.g. Fliegl\cite{fliegl2005coupled, fliegl2006inclusion}, Tew\cite{tew2007quintuple} or Ten-no.\cite{ten2004explicitly}
 Shiozaki\cite{shiozaki2008explicitly} presented the full form of the CC method up to the quadruple 
 exctitaions for ground state (CC-R12), excited states (EOM-CC-R12) and for the $\Lambda$ equation ($\Lambda$-CC-R12) of the CC analytical gradient theory.
 
 In this work we propose introducing the explicitly correlated wavefunction to the computation of the one electron density matrix (1RDM) and the cumulant of the two-electron density matrix ($\Lambda$-2RDM) in the framework of the expectation value coupled cluster theory (XCC)\cite{jeziorski1993explicitly, korona2006one, korona2008two}. In this way we propose more accurate method to the computation of one- and two-electron properties of the ground state, while making use of the XCC ability of highly controllable approximations, at relatively low cost.

\section{The CC-F12 theory}
In the CC-F12 theory the wavefunction $\Psi_0$ is represented by the usual coupled cluster
expansion
\equl{\Psi_0 = e^T\Phi_0}{ansatz}
where $\Phi_0$ is the reference determinant usually Hartree-Fock determinant, 
 and
the cluster operator $T$ is a sum of $n$-tuple excitation operators \cite{}
\equ{
  T = \sum_{n=1}^N T_n
}
where $N$ is the number of electrons.
Each of the cluster operators can be represented by the product of singlet excitation operators $E_{ai}$\cite{paldus1988clifford}
\equ{T_n = \frac{1}{n!}\sum_{\mu_n}^{N} t_{\mu_n}\mu_n = \frac{1}{n!}\sum_{\mu_n}^{N} t_{\mu_n}E_{ai}E_{bj}\ldots E_{fm},
}
where $\mu_n$ denotes $n$-th excitation level.
The indices $a, b, c \ldots$, $i, j, k \ldots$ and $p, q, r \ldots$ denote virtual, occupied and general orbitals, respectively,see \Frt{}.
When we restrict the excitations to single and double the cluster operator is composed of the standard part supplemented by the explicitly correlated component,
\equsl{&T = T_1 + T_2 + T_2', \\&T_2' = \frac12\sum_{ijkl}(t_2')^{kl}_{ij}\left[
    \sum_{\alpha\beta}\brakett{\alpha\beta}{f_{12}}{kl}E_{\alpha i}E_{\beta j} \right.
    \\
    &\left.- \sum_{ab}\brakett{ab}{f_{12}}{kl}E_{a i}E_{b j}\right]
    }{rownanie-t}
where $\alpha, \beta \ldots$ denote the complete set of orbitals, $f_{12}$
is the $r_{12}$-dependent correlation factor. The new operator $T_2'$ should satisfy 
the condition
\equ{T_2' = \hat{Q}_{12}T_2'
}
in order to assure that $T_2'$ is strongly orthogonal to products of occupied orbitals. This ensures that $T_2$ produces only two-electron correlation effect.
The $\hat{Q}_{12}$ can take several forms, among which is the so-called ansatz-3 proposed by Veleev\cite{valeev2004improving}
\equm{\hat{Q}_{12} = (1-\hat{O}_{1})(1-\hat{O}_{2}) - \hat{V}_1\hat{V}_2\\
=\hat{V}_{1}(1-\hat{P}_{2}) + (1-\hat{P}_{1})\hat{V}_{2} + (1-\hat{P}_{1})(1-\hat{P}_{2})
}
where $\hat{O}_{i}$,  $\hat{V}_{i}$ and $\hat{P}_{i}$ are the projections onto the occupied, virtual, and all orbital basis orbitals respectively, and $(1-\hat{P}_{i})$ projects  on the set of virtual orbitals of the complete basis that does not include virtual orbitals from orbital basis, see \Frt{spaces} and \Frf{ind}. This particular form of the $\hat{Q}_{12}$ operator allows us to approximate the $(\hat{1}-\hat{P})$ subspace instead of approximating the whole space $\hat{1}$.

\begin{figure}
    \centering
    \includegraphics[width= 3cm]{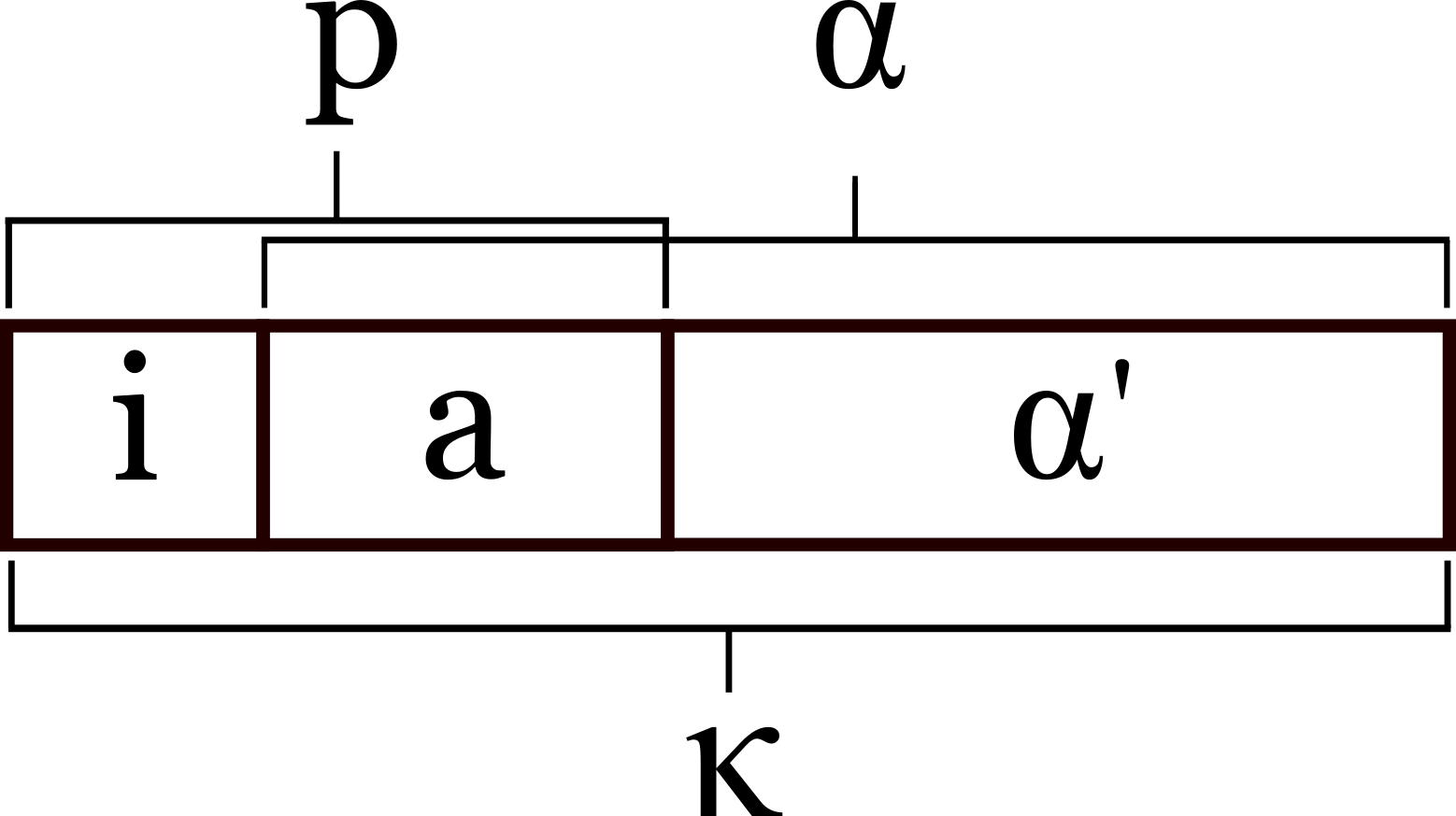}
    \caption{Partition of orbital space in CABS R12 with corresponding indices}
    \label{ind}
\end{figure}

\begin{table}\label{spaces}
    \centering
    \resizebox{\columnwidth}{!}{%
    \begin{tabular}{lll}
    \hline
       $\hat{O}_{i}$  & $i, j, k, l\ldots$&occupied \\
       $\hat{V}_{i}$  & $a,b,c,d\ldots$ & virtual in OBS\\
       $\hat{P}_{i}$  & $p,q,r,s\ldots$ &general in OBS\\
       $\hat{1}$  & $\kappa, \lambda, \mu, \nu\ldots$ & general in complete\\
       $\hat{1}-\hat{O}$  & $\alpha, \beta, \gamma, \delta\ldots$ & virtual in complete\\
       $\hat{1}-\hat{P}$  & $\alpha', \beta', \gamma', \delta'\ldots$ &virtual in complete - OBS\\
       $\hat{P}_{i}^{'}$  & $a', b', c', d'\ldots$& virtual in CABS\\
       \hline
    \end{tabular}}
    \caption{Projectors on spaces and corresponding indices}
    \label{tab:indeksy}
\end{table}

The operator $T_2'$ is a product of the geminal amplitudes $(t_2')^{kl}_{ij}$ and molecular integrals $F^{\alpha\beta}_{kl}$ involving
an explicit $r_{12}$ dependent factor.
\equa{&F^{\alpha\beta}_{kl} = \int\int d\mathbf{r}_1d\mathbf{r}_1\phi_\alpha(\mathbf{r}_1)^*\\
 & \phi_\beta(\mathbf{r}_2)^* f_{12}(\phi_k(\mathbf{r}_1)\phi_l(\mathbf{r}_1)-\phi_l(\mathbf{r}_1)\phi_k(\mathbf{r}_1))
}
with $\hat{P}_1\phi_\alpha=0$ or $\hat{P}_2\phi_\beta=0$, and $F^{\alpha\beta}_{kl}=0$ otherwise. For the  $f_{12}$ correlation factor we used
the Slater-type function of $r_{12}$
\equ{f_{12} = (-\gamma r_{12}).
}
The Ansatz \fr{ansatz}, with thus defined $T$ amplitudes, \fr{rownanie-t}, is incorporated into the Schrodinger equation 
\equ{\hat{H}\Psi_0=E\Psi_0}
with Hamiltonian $\hat{H}$ defined as
\equ{\hat{H} = \sum_{\kappa\lambda} h_{\kappa\lambda} E_{\kappa\lambda} + \frac12 \sum_{\kappa\lambda\zeta\tau}g_{\kappa\lambda\zeta\tau}E^{\kappa\lambda}_{\zeta\tau},
}
where $\kappa,\lambda,\zeta\ldots$ denote general indices in a complete basis, \Frt{tab:indeksy}.
The equation is then 
multiplied from the left by $e^{-T}$ and projected into the excited manifold 
producing the  full CCSD-F12 expression for the energy, and CC-F12 amplitudes
\equa{&\brakett{\Phi_0}{\bar{H}}{\Phi_0}\\
  &\brakett{\Phi^a_i}{\bar{H}}{\Phi_0}\\
  &\brakett{\Phi^{ab}_{ij}}{\bar{H}}{\Phi_0}\\
  &\brakett{\Phi^{kl}_{ij}}{\bar{H}}{\Phi_0}.}
$\bar{H}$ denotes the similarity transformed Hamiltonian $e^{-T}\hat{H}e^T$.

\section{Elimination of the complete basis set}
In the CC-F12 theory, the elimination of indices $\kappa, \lambda\ldots$ which run through the complete basis set is necessary in order to obtain expressions that are computationally manageable.  \Frt{tab:indeksy} and \Frf{ind} summarize how the space is divided in the CABS approach, and associates the specified indices with the corresponding subspace. The choice of the operator $\hat{Q}_{12}$ allows for an efficient approximation of the special intermediates of the F12 theory
\equsl{& \mathcal{V}^{pq}_{ij} = \frac12 g_{pq\alpha\beta}F^{\alpha\beta}_{ij}\\
  & \mathcal{X}^{kl}_{ij} = \frac12 F_{\alpha\beta}^{kl}F^{\alpha\beta}_{ij}\\
  & \mathcal{B}^{kl}_{ij} = \frac12 F_{\alpha\beta}^{kl}f_{\alpha\gamma}F^{\beta\gamma}_{ij}\\
  &  \mathcal{P}^{kl}_{ij} = \frac12 F_{\alpha\beta}^{kl}f_{\alpha\beta\gamma\delta}F^{\gamma\delta}_{ij}
}{spec}
which are rewritten in terms of the products of two-electron integrals expressed in either the OBS basis or the complete basis  belonging to the $\hat{1}-\hat{P}$ subspace. 
The complete basis is further approximated by the finite CABS basis belonging to the $\hat{P}^{'}$ subspace
\equ{\phi_{\alpha'}\approx\phi_{a'}.
}
The special intermediates are identified in the orbital expressions, before any approximations take place, and are marked
for an evaluation of an external integral engine during the computation stage.
All other terms involving the summation over the complete basis are approximated by replacing
$\alpha'$ by $a'$.



\section{XCC approach to the computation of properties}
In the literature, there are several rigorous approaches that can be extended  to calculate the molecular properties of the CC-F12 theory. The first approach base on the differentiation of CC energy expressions was introduced by Monkhorst\cite{monkhorst1977calculation, dalgaard1983some} in 1997 and later extended by Bartlett\cite{jorgensen2012geometrical, fitzgerald1986analytic, salter1987property} et. al. and is known as the $\Lambda$ vector technique. Koch and
Jørgensen\cite{jo1988mo, helgaker1989configuration, koch1990coupled} 
proposed the time-averaged quasi-energy Lagrangian technique TD-CC.
In these approaches a set of linear response equations must be solved to obtain the $\Lambda$ vector.
With this quantity at hand the CC expectation value can be calculated from a non-symmetric
expression alike
\equ{\bar{X} = \braket{(1+\Lambda)e^{(-T)}Xe^T}, 
}
where $\braket{A}$ denotes the expectation value of an operator A with a reference wavefunction, $\Phi_0$.

The second approach called XCC theory is based on the computation of molecular properties directly from the average value of an operator\cite{jeziorski1993explicitly, korona2006one} 
\equ{\bar{X}=\frac{\brakett{\Psi_0}{X}{\Psi_0}}{\braketd{\Psi_0}{\Psi_0}}.}
The wavefunction $\Psi_0$ is parameterized by the CC ansatz, and an auxiliary operator S is introduced by means of the following formula
\equl{e^S\Phi_0 = \frac{e^{T^\dagger}e^T\Phi_0}{\braketd{e^T}{e^T}}\qquad S=S_1+S_2+\ldots S_N
}{s-def}
where N is the number of electrons in the system. With help of this auxiliary operator the CC
expectation value is rewritten as
\equ{\bar{X}=\braket{e^{S^\dagger}e^{-T}Xe^Te^{-S^\dagger}}.
}
The average value of X can then be expressed as a finite series of commutators in this approach.

It is important to differentiate the XCC method discussed in this work from the approach developed by Bartlett and Noga with the same name.\cite{bartlett1988expectation} Also the operator $S$ was introduced by 
Arponen and coworkers in the context
of the extended coupled cluster theory (ECC).\cite{arponen1983variational, arponen1984variational, arponen1987extended}
However, the $S$ operator was defined in their work by a set of nonlinear equations for which no systematic approximation scheme existed. Later, Jeziorski and Moszynski\cite{jeziorski1993explicitly} proposed an expression for $S$ that could be systematically approximated by satisfying a set of linear equations.
The main finding of their work describing the $S$ operator technique is that the operator
$S$ is related to the operator $T$ by means of a relatively simple linear equation. Moreover,
this equation does not need to be solved in practice. In fact, the operator $S$  can be expanded
in a combined power series of the cluster operators $T$ and $T^\dagger$,
\equml{ S_n = T_n  -\frac{1}{n}\PP_n[\sum_{k=1}\sum_{p=1}\frac{1}{k!}p[T_p^\dagger, T]_k \\
+\sum_{k=1}\sum_{m=0}\sum_{p=1}\frac{1}{k!}
  \frac{1}{m!}p[[S_p, T^\dagger]_k,T]_m]
}{rown-s}
and $[A,B]_k$ is a shorthand for a $k$-times nested commutator. The superoperator $\hat{\mathcal{P}}_n(X)$
 yields the excitation part of $X$
 \equl{\hat{\mathcal{P}}_n(X) = \frac{1}{n!}\sum_{\mu_n} \braket{\mu_n|X}\mu_n.}{superp}
where for simplicity we introduce the following notation $\braket{A|B} = \braket{A\Phi_0|B\Phi_0}$. It is clear
from the expressions above that it can be truncated, e.g. on the basis of the perturbation
theory arguments. This also constitutes the biggest advantage of the $S$ operator technique over
the $\Lambda$ method. The task of solving the response equations to obtain $\Lambda$ is almost as expensive
as the original CC iterations themselves. Computation of the operator $S$, on the other hand,
is a relatively simple one-step (non-iterative) procedure which can be accomplished very efficiently. Since its introduction, the $S$ operator technique has been applied to calculation of the
molecular properties at the conventional CCSD and CC3 levels\cite{tucholska2014transition}, CC3 transition moments
between the ground and excited states\cite{tucholska2014transition} and excited to excited states\cite{tucholska2017transition}, electrostatic and
exchange contributions to the interaction energies of closed-shell systems\cite{moszynski1994many,korona2006one, korona2006time} and others. The
$S$ operator technique has, not yet been utilized in the context of explicitly correlated wavefunctions.


\section{F-12 expressions for the Operator $\mathbf{S}$}
In the explicitly correlated version of CC, the $T_2$ amplitudes are supplemented by the explicitly correlated component from \fr{rownanie-t}. For the purpose of deriving expressions for the $S$ amplitudes, we rewrite the T amplitudes in a more compact form
\equsl{&T_1 = \sum_{\alpha i}t^{\alpha}_i h(\alpha) E_{\alpha i}\\
&T_2 = \frac12\sum_{\alpha\beta ij}\left(\bar{\bar{t}}^{\alpha\beta}_{ij}p(\alpha)p(\beta)
+t^{\alpha\beta}_{ij}h(\alpha)h(\beta)\right) E_{\alpha i} E_{\beta j}\\
}{t-compactc}
where
\equ{\bar{\bar{t}}^{\alpha\beta}_{ij} = \sum_{kl}(t_2^{'})^{kl}_{ij}F^{\alpha\beta}_{kl}}
and $h(\alpha)$ and $p(\alpha)$ are defined as
\equs{
&h(\alpha) = 0 \quad \mbox{if}\quad \alpha \in (1-\hat{P})\\
&h(\alpha) = 1 \quad \mbox{if}\quad \alpha \in \hat{V}\\
&p(\alpha) = 1 \quad \mbox{if}\quad \alpha \in (1-\hat{P})\\
&p(\alpha) = 0 \quad \mbox{if}\quad \alpha \in \hat{V}.
}
Expressing the $T$ amplitudes in a complete basis allows us to re-derive the expressions for the $S$ amplitudes starting from \fr{s-def}
\equl{e^{T^\dagger}e^T\Phi_0 = \braketd{e^T}{e^T}\Phi_0.
}{s-mn}
We will not follow the full derivation as it can be found in the original work \cite{jeziorski1993explicitly},
instead we only note changes necessary to obtain the $S$-F12 amplitudes. 

We act on both sides of \fr{s-mn} with the $Q$ operator expressed in the complete basis
\equ{Q = \alpha^\dagger\alpha
}
to ensure it satisfies $[Q, T_n] = nT_n$ and  $[Q, S_n] = nS_n$ with the F12 amplitudes. Next we multiply both sides by $e^{-T}e^{-T^\dagger}e^{S}$ 
\equl{e^{-T}e^{-T^\dagger}e^{S} Q e^{-S}e^{T^\dagger}e^{T} = 0.
}{s-mul}
In order to obtain the $S$-F12 version of the set of linear equations from \fr{rown-s}, we project \fr{s-mul} onto the $n$-tuply excited states in a complete basis, to retain information on $T_2^{'}$. Only in this way we are able to recover the first approximation to the $S$ amplitudes, which should be equal to $T_2+T_2^{'}$. 
This implicates the form of the projection operator $\hat{\PP}_n$ which  spans over the complete basis, i.e.
\equal{&\hat{\mathcal{P}}_n(X) = \frac{1}{n!}\sum_{\substack{i_1\ldots i_n\\\alpha_i\ldots\alpha_n}} \braket{\mu_{\substack{i_1\ldots i_n\\\alpha_i\ldots\alpha_n}}|X}\mu_{\substack{i_1\ldots i_n\\\alpha_i\ldots\alpha_n}}
  .}{superp-comp}
  

\Fr{rown-s} is a linear equation that can be solved iteratively 
but it was proven more practical to expand $S_n$ by either MBPT 
expansion or expansion in powers of $T$. 
In this work we obtain the $T$ amplitudes from CC-F12 theory, 
where we in fact perform a summation to an infinite MBPT order. 
However, to facilitate discussion and formula verification, one should keep in mind 
that the $T$ amplitudes can be expanded into MBPT orders as follows:\cite{monkhorst1981recursive, jeziorski1993explicitly}
\equsl{&T_1^{(2)} = T_1^{\{2\}} + T_1^{\{3\}} + \ldots\\
&T_2^{(1)} = T_2^{\{1\}} + T_2^{\{2\}} + \ldots
}{pure}
where the superscript in curly braces indicates the pure MBPT order and the
superscript in round parentheses denotes the lowest MBPT order in which the term appears for the first time.
When the $T$ amplitudes are acquired from CCSD-F12 approximation,  
$T=T_1+T_2+T_2'$ the leading terms for 
the operators $S_n^{(m)}$ are
\equsl{
S_1^{(2)} &= T_1^{(2)} = T_1^{\{2\}} + T_1^{\{3\}} + \ldots\\
S_2^{(1)} &= T_2 + T_2^{'}\\
S_1^{(3)} &= \hat{\mathcal{P}}_1\left ([T_1^{\dagger}, T_2 + T_2'] \right ) \\
S_2^{(3)} &= \frac12\hat{\mathcal{P}}_2\left ([[T_2^{\dagger}
+(T_2^{\dagger})', T_2+T_2'], T_2+T_2'] \right ).\nonumber
}{przybl-s}

We stress that because of the CCSD approximation we are not including some of the
low order terms, that are expressed through $T_3$ or higher amplitudes, e.g. $S_3^{(2)} = T_3$.

The orbital expressions for the $S$ amplitudes are derived automatically by the code Paldus, developed by one of us (AT).
At this point we do not introduce any intermediates,
as the operators $S$ contain the integrals $F^{\alpha\beta}_{ij}$ expressed in a complete basis. Upon using the $S$ operator  in
computation of the properties one should first analyze the integrals and possible singularities, and only then
introduce new special intermediates and later on the CABS basis.

 \begin{table*}[!ht]
   \caption{Orbital expressions for the explicitly correlated $\hat{S}$ operators. Expressions for $(W1)_{aibj}, (W2)_{\alpha ibj}$ and $(W3)_{\alpha\beta ij}$ can be found in supplementary material. }\label{tabelas2}
   \vspace{0.2cm}
   \hrule
   \begin{alignat}{2}
     &\hat{S}_1^{(2)} = \sum_{ai}( s^{a}_{i})^{(2)}\hat{E}_{ai}&&= \sum_{ai}(t^{a}_{i})^{(2)}\hat{E}_{ai}\\
     &\hat{S}_1^{(3)} = \sum_{\alpha i} (s^{\alpha}_{i})^{(3)} \hat{E}_{\alpha i} &&=
  -\sum_{abij}(t_{ji}^{ab})^{(1)}(t_{j}^{b})^{(2)}\hat{E}_{ai}
+2\sum_{abij} (t_{ij}^{ab})^{(1)}(t_{j}^{b})^{(2)}\hat{E}_{ai}\\
&&&-
\sum_{\alpha a ijkl}(t_{il}^{jk}F_{jk}^{a\alpha })^{(1)}(t_{l}^{a})^{(2)}\hat{E}_{\alpha i} 
 +2\sum_{\alpha a ijkl}(t_{il}^{jk}F_{kj}^{a\alpha })^{(1)}(t_{l}^{a})^{(2)}\hat{E}_{\alpha i} \\
 &\hat{S}_2^{(1)}= \frac12\sum_{\alpha\beta ij }(s^{\alpha\beta}_{ij})^{(1)} \hat{E}_{\alpha i}\hat{E}_{\beta j}&&=\frac12
 \sum_{abij} (t^{ab}_{ij})^{(1)} \hat{E}_{ai}\hat{E}_{bj} + \frac12\sum_{\alpha\beta ijkl}(t^{kl}_{ij}F^{\alpha\beta}_{kl})^{(1)}\hat{E}_{\alpha i}\hat{E}_{\beta i}\\
  &\hat{S}_2^{(3)}= \frac12\sum_{\alpha\beta ij }(s^{\alpha\beta}_{ij})^{(3)} \hat{E}_{\alpha i}\hat{E}_{\beta j}&&=
 \frac12\sum_{abij} (W1)_{aibj} \hat{E}_{ai}\hat{E}_{bj}
 + \frac12\sum_{\alpha b ijkl}(W2)_{\alpha i bj}\hat{E}_{\alpha i}\hat{E}_{b j}\\&&&
 + \frac12\sum_{\alpha\beta ijkl}(W3)_{\alpha\beta ij}\hat{E}_{\alpha i}\hat{E}_{\beta j}
   \end{alignat}
       \hrule
 \end{table*}

\section{Density matrix in the XCC-f12 theory}
One-electron reduced density matrix (1-RDM) of an $N$-electron wavefunction $\psi$ in configuration space is defined as
\equa{\rho_{1}\left( x_{1}\right) &=N\int \psi ^{\ast }\left( x_{1},x_{2}\ldots x_{n}\right) \\&
\times\psi \left( x_{1}\ldots x_{N}\right) dx_{2}\ldots dx_{N},
}
and the average value of an arbitrary operator $\hat{X}$  can be obtained as
\equ{ \langle \hat{X}\rangle  =\int [\hat{X}\rho_1(x;x')]_{x'=x} dx= \int \hat{X}\rho(x) dx,
}
where the last equality holds for multiplicative operators.
In second quantization 1-RDM is usually denoted as $\gamma_{\kappa\lambda}$ and in a spin-adapted form is defined through singlet excitation operators $E_{\kappa\lambda}$ as
\equ{\gamma_{\kappa\lambda} = \brakett{\Psi_0}{E_{\kappa\lambda}}{\Psi_0}.}
In the case of XCC theory the density matrix is expressed with the use of the operators $S$, \fr{rown-s}.
\equ{\gamma_{\kappa\lambda}= \braket{\esd\etm E_{\kappa\lambda}\et\esdm}.
}
Because the expression for the expectation value is in a form of $e^{-Y}Xe^Y$ it is easily seen from the Baker-Campbell-Hausdorff expansion formula that it is in fact a sum of multiple commutators of connected quantities, and is therefore explicitly connected. 

For the XCCSD-F12 approximation the explicit form of this equation is
\equsl{\gamma_{\kappa\lambda} &=  \brz{E_{\kappa\lambda}} \\ 
&+ \braketd{S_1}{E_{\kappa\lambda}} + \brz{[E_{\kappa\lambda}, T_1]} + \braketd{S_2}{[E_{\kappa\lambda}, T_2+T_2']} \\ 
&+ \braketd{S_1}{[E_{\kappa\lambda}, T_2+T_2']} \\&
+ \braketd{S_1}{[E_{\kappa\lambda}, T_1]}
+ \braketd{S_2}{[[E_{\kappa\lambda}, T_1], T_2+T_2']}\\& 
+\frac12\braketd{S_1^2}{[E_{\kappa\lambda}, T_2+T_2']}\\
&+\frac12\braketd{S_1S_2}{[[E_{\kappa\lambda}, T_2+T_2'], T_2+T_2']}  \\ 
&+\frac12\braketd{S_1}{[[E_{\kappa\lambda}, T_1], T_1]}\\&+ \frac12\braketd{S_3}{[[E_{\kappa\lambda}, T_2+T_2'], T_2+T_2']}\\ 
&+\frac12\braketd{S_1^2}{[[E_{\kappa\lambda}, T_1], T_2+T_2']}\\ 
&+\frac{1}{12}\braketd{S_1^3}{[[E_{\kappa\lambda}, T_2+T_2'], T_2+T_2']}. 
}{onerdm}
This is a complete expression within the CCSD-F12 approximation. The term $S_3$ appearing in one of the terms refers to the 
$S_3^{(4)} = \hat{\mathcal{P}}_3\left ([[T_1^{\dagger}, T_2 + T_2'],  T_2 + T_2'] \right )$ which is of leading 4th MBPT order. We do not consider
the $S_3^{(4)}$ in this work. The overall leading order of this term is 5. 

Because of the absence of $T_3$ amplitudes, some of the low MBPT order terms 
are not included in the $\gamma_{\kappa\lambda}$. Specifically the term  $\braketd{S_2}{[X, T_3]}$ 
of leading $3$rd order is absent. 
Therefore, overall the XCCSD-F12 expression for 1-RDM is correct through the $2$nd MBPT order. 

The expression for the expectation value  is dependent on the operator used and the characteristics of the special intermediates
$\mathcal{Z}_{ijkl} = F^{ij}_{\gamma\alpha }x_{\alpha\beta }F_{kl}^{\beta\gamma }$ that arise in the calculation of the average value of an operator due to the presence of the $F^{\alpha\beta}_{kl}$ integrals.

In the subsequent sections, we will derive the expectation value of a general operator $\hat{X}$ in the complete basis, \frs{x-in-compl}, without making any assumptions about the nature of the special intermediates. In \frs{x-in-cabs} we assume that the CABS basis can be introduced prior to performing the multiplication of the special intermediates, and derive the corresponding expressions for the density matrix.


\subsection{Expression for the expectation value with complete indices}\label{x-in-compl}
The expression for the average value of an operator in the complete basis can be rewritten as
\equs{
  &\hat{X} = \sum_{\kappa\lambda}x_{\kappa\lambda}\hat{E}_{\kappa\lambda} \\&=
  \sum_{\alpha i} x_{\alpha i} \hat{E}_{\alpha i}  
  +   \sum_{\alpha i} x_{i \alpha} \hat{E}_{i \alpha}\\&
  + \sum_{\alpha\beta} x_{\alpha\beta}\hat{E}_{\alpha\beta}
    + \sum_{ij} x_{ij}\hat{E}_{ij}
}
From \fr{onerdm} we take only terms that are quadratic in $T$  and within
 we write only nonzero contributions
 All of the terms are summarized in \Frt{tabcomp}. In this expression we identify the special
 intermediate defined in the preceding sections $\mathcal{X}^{kl}_{ij}$ and  $\mathcal{Z}_{ijkl} $.
\begin{table*}[!ht]
  \caption{XCCSD-F12 expression for the expectation value of an operator.
    Only terms up to quadratic in $T$ are taken. }\label{tabcomp}
  \renewcommand{\arraystretch}{2.5}
  \centering 
  \begin{threeparttable}
    \vspace{0.2cm}
    \hrule    
    \begin{alignat}{2}
      &\sum_{ij}x_{ij}\braket{\hat{E}}_{ij} &&=2\sum_ix_{ii}
\\&
\sum_{ij}x_{ij} \braketd{S_1^{(2)}}{[\hat{E}_{ij}, T_1]}&&=-2\sum_{aij}x_{ij}t_{i}^{a}t_{j}^{a}\\
&
\sum_{ij} x_{ij}\braketd{S_2^{(1)}}{[\hat{E}_{ij}, T_2+T_2']}
&&= 2\sum_{abkij}x_{ij}t_{ik}^{ab}t_{jk}^{ba}
-4\sum_{abkij}x_{ij}t_{ik}^{ab}t_{jk}^{ab}\\&&&
+\sum_{\substack{ijokl\\mn \alpha\beta}}x_{ij}t_{jl}^{mn}\mathcal{X}^{mn}_{ko} (2t_{il}^{ok}-4t_{il}^{ko})\\&
\sum_{i\alpha }x_{i\alpha}\brz{[\hat{E}_{i\alpha}, T_1]} &&= 2\sum_{ai}x_{ia}t_{i}^{a}\\
&
\sum_{i\alpha}x_{i\alpha}\braketd{S_1^{(2)}}{[\hat{E}_{i\alpha}, T_2+T_2']}
&&=-2\sum_{abj}x_{ia}t_{ji}^{ab}t_{j}^{b}
+4\sum_{abj}x_{ia}t_{ij}^{ab}t_{j}^{b}\\&&&
+\sum_{ajkl}x_{i\alpha }F_{jk}^{a\alpha }t_{l}^{a}(4t_{il}^{kj}-2t_{il}^{jk})
\\
      & \sum_{\alpha i}x_{\alpha i} \braketd{S_1^{(2)}+S_1^{(3)}}{\hat{E}_{\alpha i}}&& =
2\sum_{a i} x_{ai}t_{i}^{a}
-2\sum_{abij}x_{ai}t_{ji}^{ab}t_{j}^{b}
      +4\sum_{abij}x_{ai}t_{ij}^{ab}t_{j}^{b}\\&&&
      +\sum_{ajkl}x_{\alpha i}F_{jk}^{a\alpha }t_{l}^{a}(4t_{il}^{kj}-2t_{il}^{jk})\\&
\sum_{\alpha\beta} x_{\alpha\beta}\braketd{S_2^{(1)}}{[\hat{E}_{\alpha\beta}, T_2+T_2']}
&&=-2\sum_{abcij}x_{ab}t_{ij}^{ac}t_{ji}^{bc}
+4\sum_{abcij}x_{ab}t_{ij}^{ac}t_{ij}^{bc}
\\&&&
+ 4\sum_{abijkl}x_{\alpha b}F_{ji}^{a\alpha }t_{lk}^{ab} ( 2t_{kl}^{ij}-t_{lk}^{ij})\\&&&
+ \sum_{abijkl}\mathcal{Z}^{ij}_{kl} ( 4t_{nm}^{ij}t_{nm}^{kl}-2t_{nm}^{ij}t_{mn}^{kl})
\\  &
\sum_{\alpha\beta}x_{\alpha\beta} \braketd{S_1^{(2)}}{[\hat{E}_{\alpha\beta}, T_1]}
&&=2\sum_{abi}x_{ab}t_{i}^{a}t_{i}^{b}
   \end{alignat}
  \end{threeparttable}
  \end{table*}
\subsection{Expression for the density matrix in CABS basis}\label{x-in-cabs}
For the operators that does not require special treatment of the integrals   $\mathcal{Z}_{ijkl}$ it is possible to define the density matrix
$\gamma_{\kappa\lambda}$.

Because $\kappa, \lambda$ are general indices, we distinguish nine separate blocks of the density matrix
\equ{ \gamma_{ij}, \gamma_{ai}, \gamma_{ia}, \gamma_{ab}, \gamma_{iA}, \gamma_{Ai}, \gamma_{Aa}, \gamma_{aA}, \gamma_{AB}
  .}
The expression from \fr{onerdm} is finite, therefore
it is theoretically possible to include all of the terms 
in calculations. In Table X in the supplementary material we present all of the contributions to the XCCSD-F12 1-RDM.
For each contribution we write the leading MBPT order and the cost of the most expensive term.

As a practical approximation we propose to take only the terms that are quadratic in $T$. This implicates, that we only include
$S_1^{(2)}, S_1^{(3)}$ and $S_2^{(1)}$. All of the contributions for thus approximated 1-RDM are presented in \Frt{density0}.
\begin{table*}[!ht]
  \caption{XCCSD-F12 expression for 1-RDM. Only terms up to quadratic in $T$ are taken. Einstein convention.}\label{all}
  \renewcommand{\arraystretch}{2.5}
  \centering 
  \begin{threeparttable}
    \vspace{0.2cm}
    \hrule    
    \begin{alignat}{2}
           &\gamma_{ij}
      =&&2\nonumber\\
     &\gamma_{ij}
       =&&2t_{ik}^{ab}t_{kj}^{ab}
-4t_{ik}^{ab}t_{jk}^{ab}
+4\tilde{\bar{t}}_{ik}^{Aa}\bar{\bar{t}}_{jk}^{Aa}
-8\tilde{\bar{t}}_{ik}^{Aa}\bar{\bar{t}}_{kj}^{Aa}
+4\tilde{t}_{ik}^{Aa}\bar{\bar{t}}_{kj}^{Aa}
-8\tilde{t}_{ik}^{Aa}\bar{\bar{t}}_{jk}^{Aa}
+2\tilde{\bar{t}}_{ik}^{AB}\bar{\bar{t}}_{jk}^{AB}\\&&&
-4\tilde{\bar{t}}_{ik}^{AB}\bar{\bar{t}}_{kj}^{AB}
+2\tilde{t}_{ik}^{AB}\bar{\bar{t}}_{kj}^{AB}
-4\tilde{t}_{ik}^{AB}\bar{\bar{t}}_{jk}^{AB}-2t_{i}^{a}t_{j}^{a}
\nonumber\\
&\gamma_{ia} =&&2t^a_i -2t_{ji}^{ab}t_{j}^{b}
+4t_{ij}^{ab}t_{j}^{b}\\
&\gamma_{ai} =&& 2(t_{i}^{a} -t_{ji}^{ab}t_{j}^{b}
     +2t_{ij}^{ab}t_{j}^{b})\\
&\gamma_{iA} =&&-4\tilde{\bar{t}}_{ij}^{Aa}t_{j}^{a}
+8\tilde{t}_{ij}^{Aa}t_{j}^{a}\\
&\gamma_{Ai} =&&2(-2\tilde{t}_{ij}^{aA}t_{j}^{a}
     +4\tilde{\bar{t}}_{ij}^{aA}t_{j}^{a})
\\
&\gamma_{aA} =&&-4\tilde{t}_{ij}^{Ab}t_{ji}^{ab}
+8\tilde{t}_{ij}^{Ab}t_{ij}^{ab}
-4\tilde{t}_{ij}^{AB}\bar{\bar{t}}_{ij}^{Ba}
+8\tilde{t}_{ij}^{AB}\bar{\bar{t}}_{ji}^{Ba}
\\
&\gamma_{Aa} =&&-2\bar{\bar{t}}_{ij}^{Ab}t_{ji}^{ab}
  +4s_{ij}^{Ab}t_{ij}^{ab}-4\bar{\bar{t}}_{ij}^{AB}\tilde{t}_{ij}^{Ba}
  +8\bar{\bar{t}}_{ij}^{AB}\tilde{\bar{t}}_{ij}^{Ba}
\\
&\gamma_{ab} =&&-2t_{ij}^{ac}t_{ji}^{bc}
  +4t_{ij}^{ac}t_{ij}^{bc}-4  \bar{\bar{t}}_{ji}^{Aa}\tilde{t}_{ij}^{Ab}
+8\bar{\bar{t}}_{ji}^{Aa}\tilde{\bar{t}}_{ij}^{Ab}+2t_{i}^{a}t_{i}^{b}
\\
&\gamma_{AB} =&&-4\bar{\bar{t}}_{ij}^{Aa}\tilde{\bar{t}}_{ij}^{Ba}
+8\bar{\bar{t}}_{ij}^{Aa}\tilde{t}_{ij}^{Ba}
-4\bar{\bar{t}}_{ij}^{AC}\tilde{\bar{t}}_{ij}^{BC}
+8\bar{\bar{t}}_{ij}^{AC}\tilde{t}_{ij}^{BC}
   \end{alignat}
  \hrule 
\end{threeparttable}
    \label{density0}
  \end{table*}
  The following symmetry should be satisfied
  \equ{\gamma_{\kappa\lambda}^{\{m\}} = \gamma_{\lambda\kappa}^{\{m\}}
  }
  where $\{m\}$ is the pure MBPT order. 
As an example we show $\gamma_{Aa}^{\{2\}}=\gamma_{aA}^{\{2\}}$.
From \Frt{density0} we take all of the terms of $\gamma_{Ai}^{\{2\}}$ and $\gamma_{iA}^{\{2\}}$
that are of $2$ leading order in MBPT, and for the $T$ amplitudes we only take their pure MBPT order 
according to \fr{pure}.
\equs{&\gamma_{aA}^{\{2\}} = -4\cdot\frac12 (F^{Ab}_{kl}t^{kl}_{ij})^{\{1\}}(t_{ji}^{ab})^{\{1\}}\\&
+8\cdot\frac12 (F^{Ab}_{kl}t^{kl}_{ij})^{\{1\}}  (t_{ij}^{ab})^{\{1\}}\\&
-4\cdot\frac12(F^{AB}_{kl}t^{kl}_{ij})^{\{1\}}  (F^{Ba}_{mn}t^{mn}_{ij})^{\{1\}}\\&
+8\cdot\frac12 (F^{AB}_{kl}t^{kl}_{ij})^{\{1\}}(F^{Ba}_{mn}t^{mn}_{ji})^{\{1\}}
}
\equs{&\gamma_{Aa}^{\{2\}} =-2 (F^{Ab}_{kl}t^{kl}_{ij})^{\{1\}} (t_{ji}^{ab})^{\{1\}} \\&
  +4 (F^{Ab}_{kl}t^{kl}_{ij})^{\{1\}} (t_{ij}^{ab})^{\{1\}} \\&
  -4\cdot\frac12 (F^{AB}_{kl}t^{kl}_{ij})^{\{1\}} (F^{Ba}_{mn}t^{mn}_{ij})^{\{1\}} \\&
  +8\cdot\frac12 (F^{AB}_{kl}t^{kl}_{ij})^{\{1\}}   (F^{Ba}_{mn}t^{mn}_{ji})^{\{1\}} 
  }

\section{Cumulants in the XCC theory with f12}
Cumulants originate from quantum field theory, and they are the analogs of the connected, size extensive part of the Green's functions.\cite{weinberg1995quantum, kutzelnigg1999cumulant}  In quantum chemistry they are formulated as the
 irreducible part of the density matrices. The 
n-RDM (where n$>$1)  can be divided into the nth order cumulant which
 is non separable, products of 1-RDMs and lower order cumulants. 
Cumulants are size extensive in contrast to the density matrices and can be consistently truncated  
which is especially important for 3-RDMs and higher. 

The cumulants in the coupled cluster framework were extensively studied by Korona\cite{korona2008two, korona2008first}. Recently this approach gathered an interest and the XCC cumulant was used in the computations of the corrections to the correlation energy in adiabatic connections approach.
\cite{cieslinski2022corrections} In this paper we present the 2-RDM cumulant in th XCC-f12 theory.
 
In second quantization in the spin-free formalism, cumulant can be written  as
\equ{\Lambda^{pr}_{qs} = \Gamma^{pr}_{qs} -\gamma_{pq}\gamma_{rs} + \frac12\gamma_{rq}\gamma_{ps}}
where $\Gamma^{pr}_{qs}$ is the two-electron reduced density matrix and can
be expressed using the singlet excitation operators $E_{pq}$ as
\equ{\Gamma^{pr}_{qs} = \brakett{0}{E_{pq}E_{rs}-\delta_{rq}E_{ps}}{0}.}
Therefore the cumulant in this formalism is 
\equa{&\Lambda^{pr}_{qs} = \brakett{0}{E_{pq}E_{rs}}{0} - \delta_{rq}\brakett{0}{E_{ps}}{0} \\
&-\brakett{0}{E_{pq}}{0}\brakett{0}{E_{rs}}{0}
  +\frac12\brakett{0}{E_{rq}}{0}\brakett{0}{E_{ps}}{0}\nonumber}
Introducing the XCC parametrization we arrive at the following expression
\equs{\Lambda^{\kappa \zeta }_{\lambda \tau } &= 
\braket{\esd\etm E_{\kappa \lambda }E_{\zeta \tau }\et\esdm}\nonumber \\
&-\delta_{\lambda \zeta }\braket{\esd\etm E_{\kappa \tau }\et\esdm}\nonumber \\
&  -\braket{\esd\etm E_{\kappa \lambda }\et\esdm}\braket{\esd\etm E_{\zeta \tau }\et\esdm}\nonumber \\&
  +\frac12\braket{\esd\etm E_{\zeta \lambda }\et\esdm}\braket{\esd\etm E_{\kappa \tau }\et\esdm}\nonumber \\
    &= \braket{\esd\etm E_{\kappa \lambda }\et\esdm\esd\et E_{\zeta \tau }\et\esdm}\nonumber \\&-\delta_{\lambda \zeta }\braket{\esd\etm E_{\kappa \tau }\et\esdm}\nonumber \\&
  -\braket{\esd\etm E_{\kappa \lambda }\et\esdm}\braket{\esd\etm E_{\zeta \tau }\et\esdm}\nonumber \\&
  +\frac12\braket{\esd\etm E_{\zeta \lambda }\et\esdm}\braket{\esd\etm E_{\kappa \tau }\et\esdm}\nonumber \\
  &= \braket{\esd\etm E_{\kappa \lambda }\et\esdm}\braket{\esd\et E_{\zeta \tau }\et\esdm}\nonumber \\&
  +  \braket{\esd\etm E_{\kappa \lambda }\et\esdm\PP(\esd\et E_{\zeta \tau }\et\esdm)}
  \nonumber \\&-\delta_{\lambda \zeta }\braket{\esd\etm E_{\kappa \tau }\et\esdm}\nonumber \\&
  -\braket{\esd\etm E_{\kappa \lambda }\et\esdm}\braket{\esd\etm E_{\zeta \tau }\et\esdm}\nonumber \\&
  +\frac12\braket{\esd\etm E_{\zeta \lambda }\et\esdm}\braket{\esd\etm E_{\kappa \tau }\et\esdm}\nonumber \\
  }\equs{&= \braket{\esd\etm E_{\kappa \lambda }\et\esdm\PP(\esd\et E_{\zeta \tau }\et\esdm)}
  \nonumber \\&-\delta_{\lambda \zeta }\braket{\esd\etm E_{\kappa \tau}\et\esdm}\nonumber \\&
  +\frac12\braket{\esd\etm E_{\zeta \lambda }\et\esdm}\braket{\esd\etm E_{\kappa \tau }\et\esdm}
}
Since the cumulant represents the connected part of the reduced density matrix, the aforementioned equation should also be connected.
The
 last two terms are disconnected but they cancel out with all the
disconnected terms that arise from evaluation of the  first term.  
Therefore we can write
\footnotesize
\equal{&\Lambda^{\kappa\zeta}_{\lambda\tau}\\&
=(\braket{\esd\etm E_{\kappa\lambda}\et\esdm\PP(\esd\et E_{\zeta\tau}\et\esdm)})_C\nonumber
}{cum-con}
\normalsize
where the subscript $C$ means taking only connected terms and keeping in mind that $T$ and $S$ operators are connected. 
The following symmetries hold for the cumulant of 2-RDM
\equ{\Lambda^{\kappa\zeta}_{\lambda\tau} =\Lambda_{\kappa\zeta}^{\lambda\tau}=\Lambda^{\zeta\kappa}_{\tau\lambda} 
  }

In \Frt{cumcom} we present the commutator expression for the XCCSD-F12 cumulant, terms up to  quadratic in $T$. In \Frt{cumtab}
we present the orbital expressions for the XCCSD-F12 cumulant in complete basis.
The function $\tilde{\delta}_{a \alpha}$ gives $0$ if $\alpha \in (1-\hat{P})$ and $1$ if $\alpha \in \hat{V}$. 

\begin{table*}[!ht]
  \caption{XCCSD-F12 expression for 2-RDM Cumulant. Only terms up to quadratic in $T$ are taken.}\label{cumcom}
  \renewcommand{\arraystretch}{2.5}
  \centering 
  \begin{threeparttable}
    \vspace{0.2cm}
    \hrule
    \equs{\Lambda^{\kappa \zeta}_{\lambda \tau} =& \langle E_{\kappa \lambda }\hat{\mathcal{P}}_1
    \left(E_{\zeta \tau }\right)  \rangle 
      + \langle E_{\kappa \lambda }\hat{\mathcal{P}}_1
    \left([E_{\zeta \tau }, T_2]\right)  \rangle 
+\langle E_{\kappa \lambda }\hat{\mathcal{P}}_1
    \left([E_{\zeta \tau }, T^{'}_2]\right)  \rangle 
+ \langle [S^{\dagger}_2, E_{\kappa \lambda }]\hat{\mathcal{P}}_1
    \left(E_{\zeta \tau }\right)  \rangle \\
      &+ \langle E_{\kappa \lambda }\hat{\mathcal{P}}_1
    \left([E_{\zeta \tau }, T_1]\right)  \rangle 
+\langle [S^{\dagger}_1, E_{\kappa \lambda }]\hat{\mathcal{P}}_1
    \left(E_{\zeta \tau }\right)  \rangle 
+ \langle [S^{\dagger}_2, E_{\kappa \lambda }]\hat{\mathcal{P}}_2
    \left([E_{\zeta \tau }, T_2]\right)  \rangle \\
&+\langle [S^{\dagger}_2, E_{\kappa \lambda }]\hat{\mathcal{P}}_2
    \left([E_{\zeta \tau }, T^{'}_2]\right)  \rangle 
+ \langle [S^{\dagger}_2, E_{\kappa \lambda }]\hat{\mathcal{P}}_1
    \left([E_{\zeta \tau }, T_2]\right)  \rangle 
+\langle [S^{\dagger}_2, E_{\kappa \lambda }]\hat{\mathcal{P}}_1
    \left([E_{\zeta \tau }, T^{'}_2]\right)  \rangle \\
&+ \langle [S^{\dagger}_2, [E_{\kappa \lambda }, T_2]]\hat{\mathcal{P}}_1
    \left(E_{\zeta \tau }\right)  \rangle 
+\langle [S^{\dagger}_2, [E_{\kappa \lambda }, T^{'}_2]]\hat{\mathcal{P}}_1
    \left(E_{\zeta \tau }\right)  \rangle 
+ \langle E_{\kappa \lambda }\hat{\mathcal{P}}_1
    \left([S^{\dagger}_1, [E_{\zeta \tau }, T_2]]\right)  \rangle \\
&+\langle E_{\kappa \lambda }\hat{\mathcal{P}}_1
    \left([S^{\dagger}_1, [E_{\zeta \tau }, T^{'}_2]]\right)  \rangle 
+ \langle [S^{\dagger}_1, E_{\kappa \lambda }]\hat{\mathcal{P}}_1
    \left([E_{\zeta \tau }, T_2]\right)  \rangle 
+\langle [S^{\dagger}_1, E_{\kappa \lambda }]\hat{\mathcal{P}}_1
    \left([E_{\zeta \tau }, T^{'}_2]\right)  \rangle \\
&+ \langle [S^{\dagger}_2, E_{\kappa \lambda }]\hat{\mathcal{P}}_1
    \left([E_{\zeta \tau }, T_1]\right)  \rangle 
+\langle [S^{\dagger}_2, [E_{\kappa \lambda }, T_1]]\hat{\mathcal{P}}_1
    \left(E_{\zeta \tau }\right)  \rangle 
    }
  \hrule 
\end{threeparttable}
  \end{table*}

\begin{table*}[!ht]
  \caption{XCCSD-F12 expression for 2-RDM Cumulant. Only terms up to quadratic in $T$ are taken.}\label{cumtab}
  \renewcommand{\arraystretch}{2.5}
  \centering 
  \begin{threeparttable}
    \vspace{0.2cm}
    \hrule
        \begin{alignat}{2}
     &\Lambda^{\alpha\gamma}_{\beta\delta}
       =&&-\frac{1}{2}t_{ij}^{ab}t_{ji}^{cd}\tilde{\delta}_{a\gamma }\tilde{\delta}_{b\alpha }\tilde{\delta}_{c\delta }\tilde{\delta}_{d\beta }
+t_{ij}^{ab}t_{ij}^{cd}\tilde{\delta}_{a\gamma }\tilde{\delta}_{b\alpha }\tilde{\delta}_{c\delta }\tilde{\delta}_{d\beta }
-\frac{1}{2}t_{ij}^{ab}t_{ij}^{cd}\tilde{\delta}_{a\gamma }\tilde{\delta}_{b\alpha }\tilde{\delta}_{c\beta }\tilde{\delta}_{d\delta }
+t_{ij}^{ab}t_{ji}^{cd}\tilde{\delta}_{a\gamma }\tilde{\delta}_{b\alpha }\tilde{\delta}_{c\beta }\tilde{\delta}_{d\delta }
\\&&&
-\frac{1}{2}t_{ij}^{ab}t_{ij}^{cd}\tilde{\delta}_{a\alpha }\tilde{\delta}_{b\gamma }\tilde{\delta}_{c\delta }\tilde{\delta}_{d\beta }
+t_{ij}^{ab}t_{ji}^{cd}\tilde{\delta}_{a\alpha }\tilde{\delta}_{b\gamma }\tilde{\delta}_{c\delta }\tilde{\delta}_{d\beta }
-\frac{1}{2}t_{ij}^{ab}t_{ji}^{cd}\tilde{\delta}_{a\alpha }\tilde{\delta}_{b\gamma }\tilde{\delta}_{c\beta }\tilde{\delta}_{d\delta }
+t_{ij}^{ab}t_{ij}^{cd}\tilde{\delta}_{a\alpha }\tilde{\delta}_{b\gamma }\tilde{\delta}_{c\beta }\tilde{\delta}_{d\delta }\nonumber
\\&&&
-\frac{1}{2}t_{kl}^{ab}\tilde{t}_{lk}^{\gamma \alpha }\tilde{\delta}_{a\delta }\tilde{\delta}_{b\beta }
+t_{kl}^{ab}\tilde{t}_{kl}^{\gamma \alpha }\tilde{\delta}_{a\delta }\tilde{\delta}_{b\beta }
-\frac{1}{2}t_{kl}^{ab}\tilde{t}_{kl}^{\gamma \alpha }\tilde{\delta}_{a\beta }\tilde{\delta}_{b\delta }
+t_{kl}^{ab}\tilde{t}_{lk}^{\gamma \alpha }\tilde{\delta}_{a\beta }\tilde{\delta}_{b\delta }
-\frac{1}{2}t_{kl}^{ab}\tilde{t}_{kl}^{\alpha \gamma }\tilde{\delta}_{a\delta }\tilde{\delta}_{b\beta }
+t_{kl}^{ab}\tilde{t}_{lk}^{\alpha \gamma }\tilde{\delta}_{a\delta }\tilde{\delta}_{b\beta }\nonumber\\&&&
-\frac{1}{2}t_{kl}^{ab}\tilde{t}_{lk}^{\alpha \gamma }\tilde{\delta}_{a\beta }\tilde{\delta}_{b\delta }
+t_{kl}^{ab}\tilde{t}_{kl}^{\alpha \gamma }\tilde{\delta}_{a\beta }\tilde{\delta}_{b\delta }
-\frac{1}{2}t_{kl}^{ab}\tilde{t}_{lk}^{\delta \beta }\tilde{\delta}_{a\gamma }\tilde{\delta}_{b\alpha }
+t_{kl}^{ab}\tilde{t}_{kl}^{\delta \beta }\tilde{\delta}_{a\gamma }\tilde{\delta}_{b\alpha }
-\frac{1}{2}t_{kl}^{ab}\tilde{t}_{kl}^{\beta \delta }\tilde{\delta}_{a\gamma }\tilde{\delta}_{b\alpha }
+t_{kl}^{ab}\tilde{t}_{lk}^{\beta \delta }\tilde{\delta}_{a\gamma }\tilde{\delta}_{b\alpha }\nonumber\\&&&
-\frac{1}{2}t_{kl}^{ab}\tilde{t}_{kl}^{\delta \beta }\tilde{\delta}_{a\alpha }\tilde{\delta}_{b\gamma }
+t_{kl}^{ab}\tilde{t}_{lk}^{\delta \beta }\tilde{\delta}_{a\alpha }\tilde{\delta}_{b\gamma }
-\frac{1}{2}t_{kl}^{ab}\tilde{t}_{lk}^{\beta \delta }\tilde{\delta}_{a\alpha }\tilde{\delta}_{b\gamma }
+t_{kl}^{ab}\tilde{t}_{kl}^{\beta \delta }\tilde{\delta}_{a\alpha }\tilde{\delta}_{b\gamma }
-\frac{1}{2}\tilde{t}_{mn}^{\beta \delta }\tilde{t}_{mn}^{\gamma \alpha }
+\tilde{t}_{mn}^{\beta \delta }\tilde{t}_{nm}^{\gamma \alpha }\nonumber\\&&&
-\frac{1}{2}\tilde{t}_{mn}^{\gamma \alpha }\tilde{t}_{nm}^{\delta \beta }
+\tilde{t}_{mn}^{\gamma \alpha }\tilde{t}_{mn}^{\delta \beta }
-\frac{1}{2}\tilde{t}_{mn}^{\alpha \gamma }\tilde{t}_{nm}^{\beta \delta }
+\tilde{t}_{mn}^{\alpha \gamma }\tilde{t}_{mn}^{\beta \delta }
-\frac{1}{2}\tilde{t}_{mn}^{\alpha \gamma }\tilde{t}_{mn}^{\delta \beta }
+\tilde{t}_{mn}^{\alpha \gamma }\tilde{t}_{nm}^{\delta \beta }\nonumber\\
&\Lambda^{\alpha\beta}_{ij}
       =&&-2t_{ij}^{ab}\tilde{\delta}_{a\beta }\tilde{\delta}_{b\alpha }
+4t_{ij}^{ab}\tilde{\delta}_{a\alpha }\tilde{\delta}_{b\beta }
-2\tilde{t}_{ij}^{\beta \alpha }
+4\tilde{t}_{ij}^{\alpha \beta }\\
&\Lambda^{\alpha i}_{\beta j}
       =&&2t_{ik}^{ab}t_{jk}^{bc}\tilde{\delta}_{a\beta }\tilde{\delta}_{c\alpha }
-4t_{ik}^{ab}t_{jk}^{cb}\tilde{\delta}_{a\beta }\tilde{\delta}_{c\alpha }
+2t_{ik}^{ab}t_{jk}^{ca}\tilde{\delta}_{b\beta }\tilde{\delta}_{c\alpha }
-4t_{ik}^{ab}t_{jk}^{ac}\tilde{\delta}_{b\beta }\tilde{\delta}_{c\alpha }
+2t_{ik}^{ab}\tilde{t}_{jk}^{b\alpha }\tilde{\delta}_{a\beta }\\&&&
-4t_{ik}^{ab}\tilde{\bar{t}}_{kj}^{b\alpha }\tilde{\delta}_{a\beta }
+2t_{ik}^{ab}\tilde{\bar{t}}_{kj}^{a\alpha }\tilde{\delta}_{b\beta }
-4t_{ik}^{ab}\tilde{t}_{jk}^{a\alpha }\tilde{\delta}_{b\beta }
+2t_{jm}^{ab}\tilde{\bar{t}}_{mi}^{a\beta }\tilde{\delta}_{b\alpha }
-4t_{jm}^{ab}\tilde{\bar{t}}_{mi}^{b\beta }\tilde{\delta}_{a\alpha }\nonumber\\&&&
+2t_{jm}^{ab}\tilde{t}_{im}^{b\beta }\tilde{\delta}_{a\alpha }
-4t_{jm}^{ab}\tilde{t}_{im}^{a\beta }\tilde{\delta}_{b\alpha }
+2\tilde{t}_{mj}^{\alpha \gamma }\tilde{t}_{im}^{\beta \gamma }
-4\tilde{t}_{mj}^{\gamma \alpha }\tilde{t}_{im}^{\beta \gamma }
+2\tilde{t}_{mj}^{\gamma \alpha }\tilde{t}_{im}^{\gamma \beta }
-4\tilde{t}_{mj}^{\alpha \gamma }\tilde{t}_{im}^{\gamma \beta }\\
&\Lambda^{\alpha i}_{j\beta }
=&&
+2t_{ik}^{ab}t_{jk}^{ac}\tilde{\delta}_{b\alpha }\tilde{\delta}_{c\beta }
-4t_{ik}^{ab}t_{jk}^{ca}\tilde{\delta}_{b\alpha }\tilde{\delta}_{c\beta }
-4t_{ik}^{ab}t_{jk}^{bc}\tilde{\delta}_{a\alpha }\tilde{\delta}_{c\beta }
+8t_{ik}^{ab}t_{jk}^{cb}\tilde{\delta}_{a\alpha }\tilde{\delta}_{c\beta }\\&&&
+2t_{jm}^{ab}\tilde{t}_{im}^{a\alpha }\tilde{\delta}_{b\beta }
-4t_{jm}^{ab}\tilde{t}_{im}^{b\alpha }\tilde{\delta}_{a\beta }
-4t_{jm}^{ab}\tilde{\bar{t}}_{mi}^{a\alpha }\tilde{\delta}_{b\beta }
+8t_{jm}^{ab}\tilde{\bar{t}}_{mi}^{b\alpha }\tilde{\delta}_{a\beta }\nonumber\\&&&
+2t_{ik}^{ab}\tilde{t}_{jk}^{a\beta }\tilde{\delta}_{b\alpha }
-4t_{ik}^{ab}\tilde{\bar{t}}_{kj}^{a\beta }\tilde{\delta}_{b\alpha }
-4t_{ik}^{ab}\tilde{t}_{jk}^{b\beta }\tilde{\delta}_{a\alpha }
+8t_{ik}^{ab}\tilde{\bar{t}}_{kj}^{b\beta }\tilde{\delta}_{a\alpha }\nonumber\\&&&
+2\tilde{t}_{im}^{\gamma \alpha }\tilde{t}_{mj}^{\beta \gamma }
-4\tilde{t}_{im}^{\gamma \alpha }\tilde{t}_{mj}^{\gamma \beta }
-4\tilde{t}_{im}^{\alpha \gamma }\tilde{t}_{mj}^{\beta \gamma }
+8\tilde{t}_{im}^{\alpha \gamma }\tilde{t}_{mj}^{\gamma \beta }\\
&\Lambda^{i \alpha }_{\beta j}
=&&-4t_{ik}^{ab}t_{jk}^{ca}\tilde{\delta}_{b\alpha }\tilde{\delta}_{c\beta }
+2t_{ik}^{ab}t_{jk}^{ac}\tilde{\delta}_{b\alpha }\tilde{\delta}_{c\beta }
-4t_{ik}^{ab}t_{jk}^{bc}\tilde{\delta}_{a\alpha }\tilde{\delta}_{c\beta }
+8t_{ik}^{ab}t_{jk}^{cb}\tilde{\delta}_{a\alpha }\tilde{\delta}_{c\beta }\\&&&
-4t_{ik}^{ab}\tilde{\bar{t}}_{kj}^{a\beta }\tilde{\delta}_{b\alpha }
+2t_{ik}^{ab}\tilde{t}_{jk}^{a\beta }\tilde{\delta}_{b\alpha }
-4t_{ik}^{ab}\tilde{t}_{jk}^{b\beta }\tilde{\delta}_{a\alpha }
+8t_{ik}^{ab}\tilde{\bar{t}}_{kj}^{b\beta }\tilde{\delta}_{a\alpha }\nonumber\\&&&
-4t_{jm}^{ab}\tilde{t}_{im}^{b\alpha }\tilde{\delta}_{a\beta }
+2t_{jm}^{ab}\tilde{t}_{im}^{a\alpha }\tilde{\delta}_{b\beta }
-4t_{jm}^{ab}\tilde{\bar{t}}_{mi}^{a\alpha }\tilde{\delta}_{b\beta }
+8t_{jm}^{ab}\tilde{\bar{t}}_{mi}^{b\alpha }\tilde{\delta}_{a\beta }\nonumber\\&&&
+\frac{1}{2}\tilde{\bar{t}}_{lk}^{\gamma \alpha }\tilde{\bar{t}}_{lk}^{\beta \gamma }\tilde{\delta}_{ij}
-\tilde{\bar{t}}_{lk}^{\gamma \alpha }\tilde{\bar{t}}_{lk}^{\gamma \beta }\tilde{\delta}_{ij}
+\frac{1}{2}\tilde{t}_{lk}^{\alpha \gamma }\tilde{t}_{kl}^{\beta \gamma }\tilde{\delta}_{ij}
-\tilde{\bar{t}}_{lk}^{\alpha \gamma }\tilde{\bar{t}}_{lk}^{\beta \gamma }\tilde{\delta}_{ij}\nonumber\\&&&
+\frac{1}{2}\tilde{\bar{t}}_{lk}^{\alpha \gamma }\tilde{\bar{t}}_{lk}^{\gamma \beta }\tilde{\delta}_{ij}
-\tilde{t}_{lk}^{\alpha \gamma }\tilde{t}_{kl}^{\gamma \beta }\tilde{\delta}_{ij}
+\tilde{t}_{kl}^{\gamma \alpha }\tilde{t}_{kl}^{\gamma \beta }\tilde{\delta}_{ij}
-\frac{1}{2}\tilde{t}_{kl}^{\gamma \alpha }\tilde{t}_{kl}^{\beta \gamma }\tilde{\delta}_{ij}\nonumber\\&&&
-\frac{1}{2}\tilde{t}_{kl}^{\alpha \gamma }\tilde{t}_{lk}^{\beta \gamma }\tilde{\delta}_{ij}
+\tilde{t}_{kl}^{\alpha \gamma }\tilde{t}_{kl}^{\beta \gamma }\tilde{\delta}_{ij}
-\frac{1}{2}\tilde{t}_{kl}^{\alpha \gamma }\tilde{t}_{kl}^{\gamma \beta }\tilde{\delta}_{ij}
+\tilde{t}_{kl}^{\alpha \gamma }\tilde{t}_{lk}^{\gamma \beta }\tilde{\delta}_{ij}\nonumber\\&&&
-4\tilde{t}_{im}^{\gamma \alpha }\tilde{t}_{mj}^{\gamma \beta }
+2\tilde{t}_{im}^{\gamma \alpha }\tilde{t}_{mj}^{\beta \gamma }
-4\tilde{t}_{im}^{\alpha \gamma }\tilde{t}_{mj}^{\beta \gamma }
+8\tilde{t}_{im}^{\alpha \gamma }\tilde{t}_{mj}^{\gamma \beta }\nonumber\\
&\Lambda^{i \alpha }_{\beta j}
=&&+2t_{ik}^{ab}t_{jk}^{ca}\tilde{\delta}_{b\beta }\tilde{\delta}_{c\alpha }
-4t_{ik}^{ab}t_{jk}^{cb}\tilde{\delta}_{a\beta }\tilde{\delta}_{c\alpha }
+2t_{ik}^{ab}t_{jk}^{bc}\tilde{\delta}_{a\beta }\tilde{\delta}_{c\alpha }
-4t_{ik}^{ab}t_{jk}^{ac}\tilde{\delta}_{b\beta }\tilde{\delta}_{c\alpha }\\&&&
+2t_{ik}^{ab}\tilde{\bar{t}}_{kj}^{a\alpha }\tilde{\delta}_{b\beta }
-4t_{ik}^{ab}\tilde{\bar{t}}_{kj}^{b\alpha }\tilde{\delta}_{a\beta }
+2t_{ik}^{ab}\tilde{t}_{jk}^{b\alpha }\tilde{\delta}_{a\beta }
-4t_{ik}^{ab}\tilde{t}_{jk}^{a\alpha }\tilde{\delta}_{b\beta }\nonumber\\&&&
+2t_{jm}^{ab}\tilde{t}_{im}^{b\beta }\tilde{\delta}_{a\alpha }
-4t_{jm}^{ab}\tilde{\bar{t}}_{mi}^{b\beta }\tilde{\delta}_{a\alpha }
+2t_{jm}^{ab}\tilde{\bar{t}}_{mi}^{a\beta }\tilde{\delta}_{b\alpha }
-4t_{jm}^{ab}\tilde{t}_{im}^{a\beta }\tilde{\delta}_{b\alpha }\nonumber\\&&&
+2\tilde{t}_{mj}^{\gamma \alpha }\tilde{t}_{im}^{\gamma \beta }
-4\tilde{t}_{mj}^{\gamma \alpha }\tilde{t}_{im}^{\beta \gamma }
+2\tilde{t}_{mj}^{\alpha \gamma }\tilde{t}_{im}^{\beta \gamma }
-4\tilde{t}_{mj}^{\alpha \gamma }\tilde{t}_{im}^{\gamma \beta }
\nonumber
      \end{alignat}
  \hrule 
\end{threeparttable}
  \end{table*}

\begin{table*}[!ht]
  \caption{XCCSD-F12 expression for 2-RDM Cumulant continued.}
  \renewcommand{\arraystretch}{2.5}
  \centering 
  \begin{threeparttable}
    \vspace{0.2cm}
    \hrule
    \begin{alignat}{2}
      &\Lambda^{i j }_{kl}
=&&-2t_{ik}^{ab}t_{jl}^{ba}
+4t_{ik}^{ab}t_{jl}^{ab}
-2t_{ik}^{ab}\tilde{\bar{t}}_{lj}^{ab}
+4t_{ik}^{ab}\tilde{t}_{jl}^{ab}
-2t_{jl}^{ab}\tilde{\bar{t}}_{ki}^{ab}
+4t_{jl}^{ab}\tilde{t}_{ik}^{ab}
-4t_{ik}^{mn}t_{jl}^{op}X_{opnm}
+8t_{ik}^{mn}t_{jl}^{op}X_{opmn}\\&&&
+8t_{jo}^{mn}t_{ko}^{pq}X_{pqmn}\tilde{\delta}_{il}
-4t_{jo}^{mn}t_{ko}^{pq}X_{pqnm}\tilde{\delta}_{il}
+4t_{jo}^{mn}t_{ko}^{pq}X_{qpmn}\tilde{\delta}_{il}
-8t_{jo}^{mn}t_{ko}^{pq}X_{pqmn}\tilde{\delta}_{il}\nonumber\\
&\Lambda^{\alpha \gamma }_{\beta i}
=&&-2t_{ij}^{ab}t_{j}^{c}\tilde{\delta}_{a\alpha }\tilde{\delta}_{b\gamma }\tilde{\delta}_{c\beta }
+4t_{ij}^{ab}t_{j}^{c}\tilde{\delta}_{a\gamma }\tilde{\delta}_{b\alpha }\tilde{\delta}_{c\beta }
-t_{l}^{a}\tilde{\bar{t}}_{li}^{\gamma \alpha }\tilde{\delta}_{a\beta }
+2t_{l}^{a}\tilde{t}_{il}^{\gamma \alpha }\tilde{\delta}_{a\beta }
-t_{l}^{a}\tilde{t}_{il}^{\alpha \gamma }\tilde{\delta}_{a\beta }
+2t_{l}^{a}\tilde{\bar{t}}_{li}^{\alpha \gamma }\tilde{\delta}_{a\beta }\\
&\Lambda^{\alpha \beta }_{ i\gamma}
=&&-2t_{ij}^{ab}t_{j}^{c}\tilde{\delta}_{a\beta }\tilde{\delta}_{b\alpha }\tilde{\delta}_{c\gamma }
+4t_{ij}^{ab}t_{j}^{c}\tilde{\delta}_{a\alpha }\tilde{\delta}_{b\beta }\tilde{\delta}_{c\gamma }
-t_{l}^{a}\tilde{t}_{il}^{\beta \alpha }\tilde{\delta}_{a\gamma }
+2t_{l}^{a}\tilde{\bar{t}}_{li}^{\beta \alpha }\tilde{\delta}_{a\gamma }
-t_{l}^{a}\tilde{\bar{t}}_{li}^{\alpha \beta }\tilde{\delta}_{a\gamma }
+2t_{l}^{a}\tilde{t}_{il}^{\alpha \beta }\tilde{\delta}_{a\gamma }\\
&\Lambda^{\alpha j }_{ i k}
=&&+2t_{ik}^{ab}t_{j}^{a}\tilde{\delta}_{b\alpha }
-4t_{ik}^{ab}t_{j}^{b}\tilde{\delta}_{a\alpha }
+2t_{j}^{a}\tilde{t}_{ik}^{a\alpha }
-4t_{j}^{a}\tilde{\bar{t}}_{ki}^{a\alpha }\\
&\Lambda^{\alpha \beta }_{ i \gamma}
=&&-2t_{ij}^{ab}t_{j}^{c}\tilde{\delta}_{a\beta }\tilde{\delta}_{b\gamma }\tilde{\delta}_{c\alpha }
+4t_{ij}^{ab}t_{j}^{c}\tilde{\delta}_{a\gamma }\tilde{\delta}_{b\beta }\tilde{\delta}_{c\alpha }
-t_{l}^{a}\tilde{t}_{il}^{\beta \gamma }\tilde{\delta}_{a\alpha }
+2t_{l}^{a}\tilde{\bar{t}}_{li}^{\beta \gamma }\tilde{\delta}_{a\alpha }
-t_{l}^{a}\tilde{\bar{t}}_{li}^{\gamma \beta }\tilde{\delta}_{a\alpha }
+2t_{l}^{a}\tilde{t}_{il}^{\gamma \beta }\tilde{\delta}_{a\alpha }\\
&\Lambda^{i \beta }_{ \alpha \gamma}
=&&-2t_{ij}^{ab}t_{j}^{c}\tilde{\delta}_{a\gamma }\tilde{\delta}_{b\alpha }\tilde{\delta}_{c\beta }
+4t_{ij}^{ab}t_{j}^{c}\tilde{\delta}_{a\alpha }\tilde{\delta}_{b\gamma }\tilde{\delta}_{c\beta }
-t_{l}^{a}\tilde{t}_{il}^{\gamma \alpha }\tilde{\delta}_{a\beta }
+2t_{l}^{a}\tilde{\bar{t}}_{li}^{\gamma \alpha }\tilde{\delta}_{a\beta }
-t_{l}^{a}\tilde{\bar{t}}_{li}^{\alpha \gamma }\tilde{\delta}_{a\beta }
+2t_{l}^{a}\tilde{t}_{il}^{\alpha \gamma }\tilde{\delta}_{a\beta }\\
&\Lambda^{i \alpha }_{ j k}
=&&-4t_{ij}^{ab}t_{k}^{b}\tilde{\delta}_{a\alpha }
+2t_{ij}^{ab}t_{k}^{a}\tilde{\delta}_{b\alpha }
-4t_{k}^{a}\tilde{\bar{t}}_{ji}^{a\alpha }
+2t_{k}^{a}\tilde{t}_{ij}^{a\alpha }
\\
&\Lambda^{i j }_{ \alpha k}
=&&+2t_{jk}^{ab}t_{i}^{b}\tilde{\delta}_{a\alpha }
-4t_{jk}^{ab}t_{i}^{a}\tilde{\delta}_{b\alpha }
+2t_{i}^{a}\tilde{\bar{t}}_{kj}^{a\alpha }
-4t_{i}^{a}\tilde{t}_{jk}^{a\alpha }\\
&\Lambda^{i k}_{ j \alpha}
=&&+2t_{ik}^{ab}t_{j}^{b}\tilde{\delta}_{a\alpha }
-4t_{ik}^{ab}t_{j}^{a}\tilde{\delta}_{b\alpha }
+2t_{j}^{a}\tilde{\bar{t}}_{ki}^{a\alpha }
-4t_{j}^{a}\tilde{t}_{ik}^{a\alpha }
      \end{alignat}
        \hrule 
\end{threeparttable}
  \end{table*}

\section{Computational details}
The derivation of the orbital-level 
expressions in this work extremely error-prone.
We automated this process with the \paldus~code, which is designed to derive, simplify,
and automatically implement expressions of the type
\equ{\braket{[V_1,\mu_n]_{k_1}|[V_2, V_3]_{k_2}|[V_4,\nu_m]_{k_3}},
}
where $k_1, k_2, k_3$ denote $k$-tuply nested commutators. The operators $V_1-V_4$ could be any excitation, deexcitation,
or general operators that are represented by the products of the $E_{pq}$  operators.
Each of the integrals is approximated within the requested level
of theory and integrated using the Wick's theorem,\cite{wick1950evaluation}
generalized to the form of contracting and ordering $E_{pq}$ strings. This process can be a limiting step for a long $E_{pq}$ strings, especially in the F12 case, therefore 
the integration is carried out into a parallel mode. 

The result of the integration usually contain
tens of thousands of terms that need to be compared efficiently. This is done by the standardization
of each term to an unambiguous form according to index names and their permutations.  Subsequently, each
term is translated to a compiled-language representation  and the
simplification is carried out in this form, which drastically speed up the process. 

The next step after the simplification is the identification of the special intermediates
$V$, $X$, $B$, $P$, $Z$. 
%
Finally, the result is translated back and a parallel Fortran
ready to attach module is produced.

The implementation is optimized in the sense that \paldus~automatically
computes and selects the best intermediates for each term, considering
memory usage to computational time ratio. 
\section{Summary}
In this work we have presented the expressions for the 1-RDM and the 2-RDM cumulant in the framework of the XCC-F12 theory.
The reduced density matrices are quantities that are widely used in quantum chemistry. They pose an alternative to the wavefunction approach. As the density matrices similarly to wavefunctions are not extensive and can be further separated it is useful to work with the irreducible parts of the density matrices - cumulants. Cumulants are not only connected (and thus extensive) but can also be systematically approximated making them a desirable tool for demanding computations. 

In order to obtain chemical accuracy in the computations that are making use of the cumulants 
(e.g. properties) we proposed to express them in the framework of the expectation value coupled cluster theory together with the explicitly correlated wavefunction. 
In this way we obtained the expressions for 1- and 2-RMD cumulants that are based on the coupled cluster theory, are connected and can be systematically approximated. On top of that by using the explicitly correlated wavefunction we are able to obtain expressions that would generate more
accurate results at the CCSD level without introducing the costly triples amplitudes. 

We have presented the ready-to-implement expressions for the F12 $S$ amplitudes, 1-RDM and 2-RDM cumulant. We have described the technical details needed to obtain the intermediates needed to lower the computational cost. 

\section{Acknowledgment}
 This research was supported by the National Science Center (NCN) under Grant No. 2017/25/B/ST4/02698.
{\footnotesize
  \bibliography{ref}}

\end{document}